\documentclass[aps,pra,showpacs,groupedaddress,floatfix,nofootinbib]{revtex4}

\begin{document}

\title{On the exact calculation of travelling wave solutions to nonlinear evolution
equations}
\author{Francisco M. Fern\'andez}

\affiliation{INIFTA (UNLP, CCT La Plata-CONICET), Divisi\'on Qu\'imica
Te\'orica Blvd. 113 y 64 S/N, Sucursal 4, Casilla de Correo 16, 1900 La
Plata, Argentina}

\begin{abstract}
We analyse and compare three methods for the exact calculation of travelling
wave solutions to nonlinear partial differential equations. We simplify the
so called $(G^{\prime }/G)$--expansion method and apply two of those methods
to simple physical problems.
\end{abstract}

\maketitle

\section{Introduction\label{sec:Intro}}

There has recently been great interest in the application of the $(G^{\prime
}/G)$--expansion method to obtaining travelling wave solutions of some
nonlinear partial differential equations\cite{ZG08, GA09}. The method was
earlier applied to a variety of such problems\cite{WLZ08,WZL08}. There are
many mathematical recipes for that purpose\cite{ZG08, GA09}, two of them are
the exp--function method\cite{HW06} and the tanh--function method\cite{F00}.
All those prescriptions look quite similar and have also been applied to
similar toy problems. It is curious that there has not been much interest in
comparing them.

In this paper we analyze those recipes. In Sec.~\ref{sec:G_method} we
outline the $(G^{\prime }/G)$--expansion method and in Sec.~\ref
{sec:F_method} propose a simplified version of it. In Sec.~\ref
{sec:comparison} we compare the three prescriptions just mentioned and in
Sec.~\ref{sec:examples} we apply them to two simple examples. Finally, in
Sec.~\ref{sec:conclusions} we draw conclusions from the results of the
preceding sections.

\section{The $(G^{\prime }/G)$--expansion method\label{sec:G_method}}

This method was proposed to obtain analytical travelling--wave solutions to
nonlinear differential equations of the form\cite{WLZ08}
\begin{equation}
P(u,u_{t},u_{tt},u_{x},u_{xx},u_{tx},\ldots )=0  \label{eq:dif_eq}
\end{equation}
where the subscripts indicate differentiation of $u(x,t)$ with respect to
its arguments, and $P$ is a polynomial function. The $(G^{\prime }/G)$%
--expansion method has been applied to several examples of travelling waves
that are particular solutions of the form $u(x,t)=u(\xi )$, where $\xi =x-ct$
so that $u_{t}=-cu^{\prime }$ and $u_{x}=u^{\prime }$\cite{ZG08,
GA09,WLZ08,WZL08}.

One assumes that the travelling--wave solution to the differential equation (%
\ref{eq:dif_eq}) is of the form
\begin{equation}
u(\xi )=\sum_{j=0}^{m}a_{j}\left[ \frac{G^{\prime }(\xi )}{G(\xi )}\right]
^{j}  \label{eq:u_series}
\end{equation}
where the function $G(\xi )$ is a solution to
\begin{equation}
G^{\prime \prime }(\xi )+\lambda G^{\prime }(\xi )+\mu G(\xi )=0
\label{eq:dif_eq_G}
\end{equation}
It follows from
\begin{equation}
\frac{d}{d\xi }\left( \frac{G^{\prime }}{G}\right) =-\left( \frac{G^{\prime }%
}{G}\right) ^{2}-\lambda \frac{G^{\prime }}{G}-\mu  \label{dq:dif_eq_G'/G}
\end{equation}
that $d^{k}u/d\xi ^{k}$ is a polynomial function of $G^{\prime }/G$ of
degree $m+k$. For this reason, substitution of Eq.~(\ref{eq:u_series}) into
Eq.~(\ref{eq:dif_eq}) yields a polynomial function of $G^{\prime }/G$ and if
we set its coefficients equal to zero we may obtain the unknown coefficients
$a_{j}$.

\section{The $(F^{\prime }/F)$--expansion method\label{sec:F_method}}

The function $F(\xi )=e^{\lambda \xi /2}G(\xi )$ is a solution of the
differential equation $F^{\prime \prime }+(\mu -\lambda ^{2}/4)F=0$ and
satisfies
\begin{equation}
\frac{G^{\prime }}{G}=\frac{F^{\prime }}{F}-\frac{\lambda }{2}
\label{eq:G-F}
\end{equation}
Therefore the series (\ref{eq:u_series}) becomes
\begin{equation}
u(\xi )=\sum_{j=0}^{m}b_{j}\left[ \frac{F^{\prime }(\xi )}{F(\xi )}\right]
^{j}  \label{eq:u_series_F}
\end{equation}
where the coefficients $b_{j}$ absorb the parameter $\lambda $ that
therefore dissapears from the equations. We may call $\gamma =\mu -\lambda
^{2}/4$ and simply require that the function $F(\xi )$ is a solution to
\begin{equation}
F^{\prime \prime }(\xi )+\gamma F(\xi )=0  \label{eq:dif_eq_F}
\end{equation}
For completeness and comparison we show the solutions to this trivial
equation:
\begin{equation}
F(\xi )=\left\{
\begin{array}{l}
c_{1}\cos \left( \sqrt{\gamma }\xi \right) +c_{2}\sin \left( \sqrt{\gamma }%
\xi \right) ,\,\gamma >0 \\
c_{1}\cosh \left( \sqrt{-\gamma }\xi \right) +c_{2}\sinh \left( \sqrt{%
-\gamma }\xi \right) ,\,\gamma <0 \\
c_{1}+c_{2}\xi ,\,\gamma =0
\end{array}
\right.  \label{eq:F's}
\end{equation}
Notice that this expression leads to all the cases explicitly shown by Zayed
and Gepreel\cite{ZG08}. Therefore, this method is entirely equivalent to the
$(G^{\prime }/G)$--expansion method but the equation that defines the main
function is simpler. In Sec.~\ref{sec:examples} we show an application of
the $(F^{\prime }/F)$--expansion method.

It is worth noticing that the simplification proposed here also applies to
the generalized $(G^{\prime }/G)$--expansion method\cite{ZWL08} because it
is based on the same equation for $G(\xi )$.

\section{Comparison of different methods\label{sec:comparison}}

The function $w(\xi )=-F^{\prime }(\xi )/F(\xi )$ is a solution to the
Riccati equation
\begin{equation}
w^{\prime }=\gamma +w^{2}  \label{eq:dif_eq_w}
\end{equation}
proposed by Fan\cite{F00} to expand the traveling--wave solutions in the
form
\begin{equation}
u(\xi)=\sum_{j=0}^{m}a_{j}w^{j}  \label{eq:u_series_w}
\end{equation}
We appreciate that the extended tanh--function method is equivalent in
principle to the $(F^{\prime }/F)$--expansion method and thereby to the $%
(G^{\prime }/G)$--expansion method. There is a difference, however, in that
the latter two methods appear to be more flexible because they have the two
independent solutions of the generating equation built in the expansion
function through the constants $c_{1}$ and $c_{2}$.

We can rewrite the solution to Eq.~(\ref{eq:dif_eq_F}) in a different way
\begin{equation}
F(\xi )=c_{1}e^{\alpha \xi }+c_{2}e^{-\alpha \xi },\,\alpha =\sqrt{-\gamma }
\label{eq:F_exp}
\end{equation}
so that the expansion (\ref{eq:u_series_F}), and consequently also (\ref
{eq:u_series}), reduces to a ratio of two polynomial functions of $e^{\alpha
\xi }$ of degree $2m$. We thus conclude that the $(G^{\prime }/G)$%
--expansion method is a particular case of the Exp--function method\cite
{HW06} that is based on a solution of the form
\begin{equation}
u(\eta )=\frac{\sum_{j=-c}^{d}a_{j}\exp (j\eta )}{\sum_{j=-p}^{q}b_{j}\exp
(j\eta )},\,\eta =kx+\omega t  \label{eq:u_rat}
\end{equation}
Notice that $\alpha \xi =\eta $ if $\alpha =k$ and $c=-\omega /k$. It is
clear that we can rewrite the rational approximation (\ref{eq:u_rat}) in
such a way that it only shows positive powers of $e^{\eta }$ in the
numerator and denominator.

In the following section we apply the Exp--function method to a problem that
apparently cannot be treated by means of the $(G^{\prime }/G)$--expansion
method.

\section{Applications\label{sec:examples}}

It is not the purpose of this paper to abound with arbitrary tailor--made
toy model equations for the application of the methods just described.
However, in what follows we consider two well--known exactly solvable
problems that will enable us to compare the methods outlined above.

Our firs example is Fisher's equation
\begin{equation}
\frac{\partial u}{\partial t}=\frac{\partial ^{2}u}{\partial x^{2}}+u(1-u)
\label{eq:Fisher}
\end{equation}
that was originally derived for the simulation of propagation of a gene in a
population and also arises in heat and mass transfer, biology, and ecology.
This equation is so popular that appears in most on--line encyclopaedias of
mathematics\cite{Feq}. For our present purposes it suffices to mention the
application of the Exp--function method\cite{Z08}.

If we choose $\xi =x-ct$ (we may also choose $\xi =x+ct$) we obtain
\begin{equation}
u^{\prime \prime }+cu^{\prime }+u(1-u)  \label{eq:Fisher_TW}
\end{equation}
The leading terms of $u^{\prime \prime }$ and and $u^{2}$ are of dgree $m+2$
and $2m$, respectively, and therefore they do not cancel unless $m=2$. If we
substitute Eq.~(\ref{eq:u_series_F}) into Eq.~(\ref{eq:Fisher_TW}) we easily
obtain $b_{2}=6$, $b_{1}=-6c/5$, $b_{0}=[25(8\gamma +1)-c^{2}]/50$, $\gamma
=-c^{2}/100$. The constant term gives us the only values of $c$ for which
there are exact solutions: $c=\pm 5/\sqrt{6}$. For $c=5/\sqrt{6}$ we obtain
\begin{equation}
u(\xi )=\frac{1}{\left[ 1+Ce^{\xi /\sqrt{6}}\right] ^{2}}
\label{eq:u_Fisher}
\end{equation}

By means of the Exp--function method Zhou\cite{Z08} identified four cases
that produced four solutions named $u_{j}(x,t)$, $j=1,2,3,4$. However, they
are not essentially different as one can show that they are closely related
by the symmetry of the problem: $u_{1}(-x,t)=u_{2}(x,t)$, $%
u_{3}(-x,t)=u_{4}(x,t)$, and $u_{1}(x,-t)$ becomes $u_{3}(x,t)$ after
substitution of $4/b_{0}$ for $b_{0}$ in the latter. On the other hand, Eq.~(%
\ref{eq:u_Fisher}) is identical to $u_{2}(x,t)$ if we simply substitute $%
2/b_{0}$ for $C$. The other solutions follow from the indicated symmetry or
by choosing the other root $c=-5/\sqrt{6}$.

Our second example is the Bratu--Gelfand equation
\begin{equation}
u^{\prime \prime }(x)+\lambda e^{u(x)}=0,\,u^{\prime }(0)=u(1)=0
\label{eq:Bratu}
\end{equation}
that appears in simple models for the stationary theory of the thermal
explosion\cite{F-K55}. First, we have to convert this strongly nonlinear
equation into a polynomial one, which we do by means of the transformation $%
u(x)=-n\ln v(x)$ that leads to $\lambda v^{2-n}+n\left( v^{\prime
2}-vv^{\prime \prime }\right) =0$. In order to have the simplest equation we
choose $n=2$, so that
\begin{equation}
2\left( v^{\prime 2}-vv^{\prime \prime }\right) +\lambda =0,\,v^{\prime
}(0)=0,\,v(1)=1  \label{eq:Bratu2}
\end{equation}
If we substitute
\begin{equation}
v(x)=a_{-1}e^{-\alpha x}+a_{0}+a_{1}e^{\alpha x}  \label{eq:Bratu_v}
\end{equation}
which is a particular case of Eq.~(\ref{eq:u_rat}), into Eq.~(\ref{eq:Bratu2}%
) we easily obtain
\begin{equation}
a_{0}=0,\,a_{-1}=\frac{\lambda }{8a_{1}\alpha ^{2}}  \label{eq:Bratu_a0,a1}
\end{equation}
The boundary condtions $v^{\prime }(0)=0$ and $v(1)=1$ give us two
additional equations
\begin{eqnarray}
a_{1}\alpha -\frac{\lambda }{8a_{1}\alpha } &=&0  \nonumber \\
a_{1}e^{\alpha }+\frac{\lambda e^{-\alpha }}{8a_{1}\alpha ^{2}} &=&1
\label{eq:Bratu_BC}
\end{eqnarray}
respectively, from which we obtain
\begin{eqnarray}
\lambda &=&\frac{8\alpha ^{2}e^{2\alpha }}{\left( e^{2\alpha }+1\right) ^{2}}
\nonumber \\
a_{1} &=&\frac{e^{\alpha }}{e^{2\alpha }+1}  \label{eq:Bratu_lam,a1}
\end{eqnarray}
The first equation shows a well--known bifurcation diagram from which one
obtains the critical condition of ignition $\alpha _{c}=1.19967864$ that is
a root of $d\lambda /d\alpha =0$\cite{F-K55}. In other words, explosion
takes place in a self--ignition process for a plane--parallel vessel when $%
\lambda =\lambda _{c}=\lambda (\alpha _{c})=0.8784576797$\cite{F-K55}.

Neither the $(G^{\prime }/G)$--expansion method\cite{ZG08, GA09,WLZ08,WZL08}
nor the tanh--function method\cite{F00} provide a function of the form (\ref
{eq:Bratu_v}) that is already given by the more flexible Exp--function method%
\cite{HW06}.

\section{Conclusions\label{sec:conclusions}}

There seems to be an ever increasing number of methods for the solution of
more or less trivial nonlinear problems; many of them are cited in the
papers by Zayed and Gepreel\cite{ZG08} and Ganji and Abdollahzadeh\cite{GA09}%
. Our analysis of three such recipes was motivated by the interest of this
journal in one of those methods. We have shown how to simplify the main
equation of the $(G^{\prime }/G)$--expansion method and have compared it
with the tanh--method and Exp--function methods. Our simple and
straightforward discussion shows that the first two ones are similar in
essence and that the third one is more general and flexible. The three of
them lead to solutions in the form of quotients of polynomial functions of
exponential ones.

We have illustrated the simplification of the $(G^{\prime }/G)$--expansion
method, named $(F^{\prime }/F)$--expansion method, by its application to the
simple and widely known Fisher's equation\cite{Feq,Z08}. Finally, we have
verified the greater flexibility of the Exp--function method with the aid of
the Bratu--Gelfand equation\cite{F-K55}.

The papers of Zayed and Gepreel\cite{ZG08} and Ganji and Abdollahzadeh\cite
{GA09} do not exhibit the amazing errors of other articles that I discussed
in the past\cite{F07,F08b,F08c,F08d,F08e,F08f}.\ However, they are standard
applications of a method developed earlier and widely applied in previous
papers by several authors\cite{ZG08,GA09} (see also the references cited
therein). It is suprising that Journal of Mathematical Physics (supposedly
devoted to original and more elaborate mathematical applications to physics)
accepts such kind of contributions. As an experiment I submitted a comment
to JMP (with the content shown above) and it was rejected on the grounds
that it did not ``contain sufficiently significant new results to warrant
its publication in JMP''. Does it seem that the Elsevier policy of
publishing poor papers for reasons other than purely scientific ones is
beginning to contaminate also JMP?

\end{document}